\documentclass[twocolumn,english,aps,prl,reprint,superscriptaddress]{revtex4-1}
\usepackage[LGR,T1]{fontenc}
\usepackage[latin9]{inputenc}
\usepackage{verbatim}
\usepackage{amstext}
\usepackage{amssymb}
\usepackage{graphicx}

\usepackage{chemformula}

\usepackage{braket}

\def\meV{\,\textrm{meV}}

\makeatletter

\ProvideTextCommand{\~}{LGR}[1]{\char126#1}

\@ifundefined{textcolor}{}
{%
 \definecolor{BLACK}{gray}{0}
 \definecolor{WHITE}{gray}{1}
 \definecolor{RED}{rgb}{1,0,0}
 \definecolor{GREEN}{rgb}{0,1,0}
 \definecolor{BLUE}{rgb}{0,0,1}
 \definecolor{CYAN}{cmyk}{1,0,0,0}
 \definecolor{MAGENTA}{cmyk}{0,1,0,0}
 \definecolor{YELLOW}{cmyk}{0,0,1,0}
}

\usepackage[english]{babel}
\def\urlprefix{}
\def\url#1{}
\addto\captionsenglish{%
}

\usepackage[none]{hyphenat}

\newcommand{\angstrom}{\mbox{\normalfont\AA}}

\makeatother

\begin{document}
\title{Using hyper-optimized tensor networks and first-principles electronic structure to simulate experimental properties of the giant \ch{\{Mn84\}} torus}
\author{Dian-Teng Chen}
\thanks{These two authors contributed equally}
\affiliation{Department of Physics, Center for Molecular Magnetic Quantum Materials
and Quantum Theory Project, University of Florida, Gainesville, Florida
32611, USA}

\author{Phillip Helms}
\thanks{These two authors contributed equally}
\affiliation{Division of Chemistry and Chemical Engineering, California Institute of Technology, Pasadena, California 91125, USA}

\author{Ashlyn R. Hale}
\affiliation{Department of Chemistry, Center for Molecular Magnetic Quantum Materials, University of Florida, Gainesville, Florida
32611, USA}
\author{Minseong Lee}
\affiliation{National High Magnetic Field Laboratory, Los Alamos National Laboratory, Los Alamos, New Mexico 87545, USA}
\author{Chenghan Li}
\affiliation{Division of Chemistry and Chemical Engineering, California Institute of Technology, Pasadena, California 91125, USA}

\author{Johnnie Gray}
\affiliation{Division of Chemistry and Chemical Engineering, California Institute of Technology, Pasadena, California 91125, USA}

\author{George Christou}
\affiliation{Department of Chemistry, Center for Molecular Magnetic Quantum Materials, University of Florida, Gainesville, Florida
32611, USA}

\author{Vivien S. Zapf}
\affiliation{National High Magnetic Field Laboratory, Los Alamos National Laboratory, Los Alamos, New Mexico 87545, USA}
\author{Garnet Kin-Lic Chan}
\email{gkc1000@gmail.com}
\affiliation{Division of Chemistry and Chemical Engineering, California Institute of Technology, Pasadena, California 91125, USA}
\author{Hai-Ping Cheng}
\email{hping@ufl.edu}
\affiliation{Department of Physics, Center for Molecular Magnetic Quantum Materials
and Quantum Theory Project, University of Florida, Gainesville, Florida
32611, USA}

\begin{abstract}
The single-molecule magnet \ch{\{Mn84\}} is a challenge to theory due to its high nuclearity. Building on our prior work which characterized the structure of the spectrum of this magnet, we directly compute two experimentally accessible observables, the field-dependent magnetization up to 75 T and the temperature-dependent heat capacity, using parameter free theory. In particular, we use first principles calculations to derive short- and long-range exchange interactions, while we compute the exact partition function of the resulting classical Potts and Ising spin models for all 84 Mn $S=2$ spins to obtain the observables. The latter computation is possible because of a simulation methodology that uses hyper-optimized tensor network contraction, borrowing from recent techniques developed to simulate quantum supremacy circuits. We also synthesize the magnet and measure its heat capacity and field-dependent magnetization. We observe good qualitative agreement between theory and experiment, identifying an unusual peak in the heat capacity in both, as well as a plateau in the magnetization. Our work also identifies some limitations of current theoretical modeling in large magnets, such as the sensitivity to small, long-range,  exchange couplings. 
\end{abstract}

\maketitle
 
\section{Introduction}

Single-molecule magnets (SMMs) have invoked fascination both because of the possibility to study finite analogues of bulk classical and quantum magnetic phenomena~\cite{christou2000single}, as well as for their potential in  information science applications~\cite{friedman1996macroscopic,thomas1996macroscopic,wernsdorfer2002spin,wernsdorfer1999quantum}.
The first SMM \ch{[Mn12O12(O2CR)16(H2O)4]} was synthesized in the early 1990s and since then, SMM's of increasing nuclearity have been made~\cite{muller1995mo154,muller1998formation,muller1999molecular,muller2002metal, brechin2002quantum, soler2001synthesis,vinslava2016molecules,tasiopoulos2004giant}. In the case of Mn-based SMM's, 
\ch{\{Mn18\}} \cite{brechin2002quantum}, \ch{\{Mn30\}} \cite{soler2001synthesis}, \ch{\{Mn70\}} \cite{vinslava2016molecules} and \ch{\{Mn84\}} \cite{tasiopoulos2004giant} have been reported, closing the gap between the largest bottom-up synthesized SMM and the smallest top-down synthesized magnetic nanoparticles \cite{batlle2002finite,hyeon2003chemical,laurent2008magnetic}. In fact, the giant \ch{\{Mn84\}} torus, first reported in 2004 \cite{tasiopoulos2004giant}, has  an external diameter of ~4.3 nm, making it larger than many top-down magnetic nanoparticles.

\begin{figure}
\includegraphics[width=1\columnwidth]{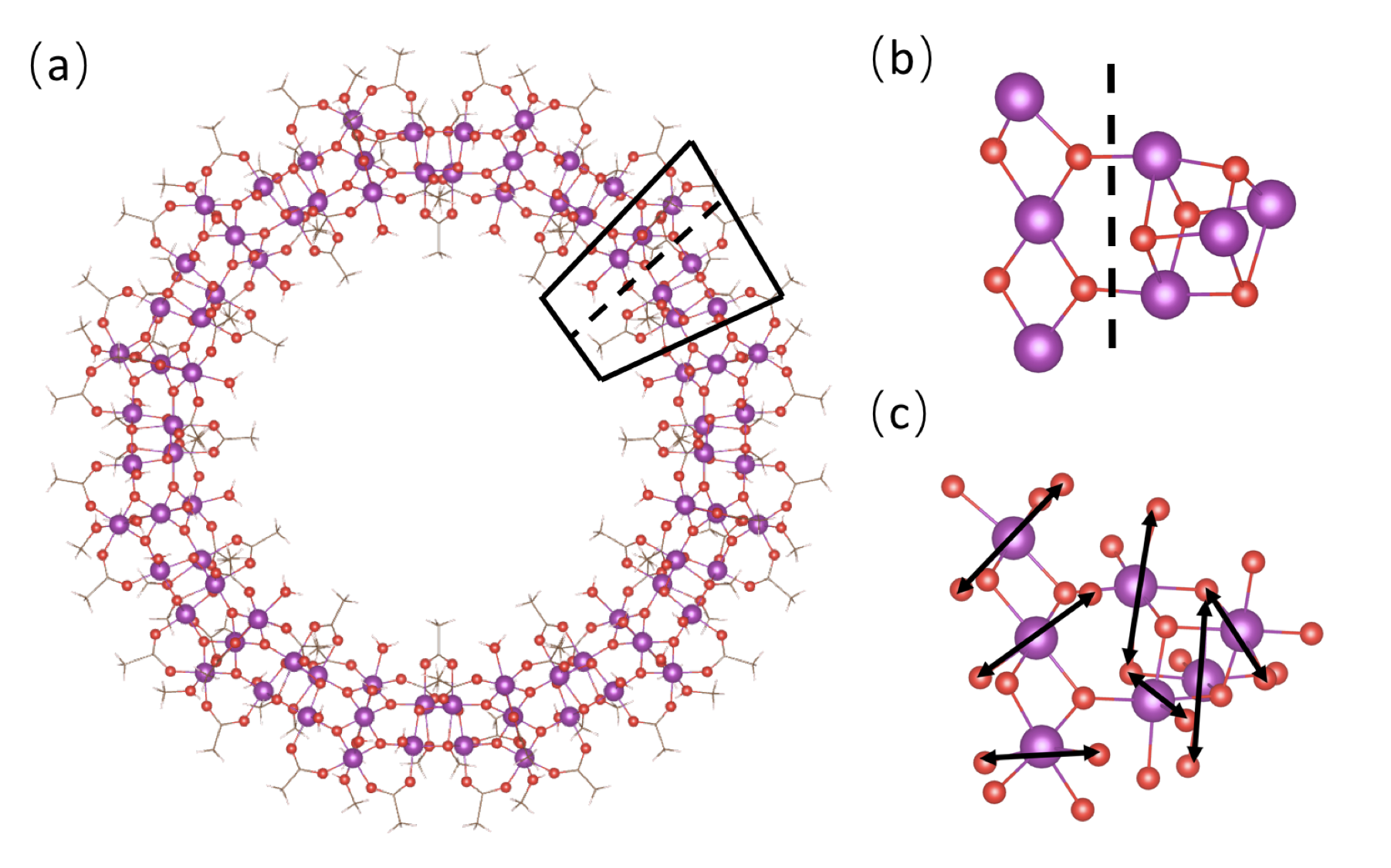}
\caption{(a) The structure of the \ch{\{Mn84\}} torus. The black box and (b) shows the alternating \ch{\{Mn3\}} (linear, left) and \ch{\{Mn4\}} (cubane, right) subunits, separated by the dashed line. (c) Jahn-Teller elongation axes for each \ch{Mn^{3+}} in the \ch{\{Mn7\}} subunit. Colour code: Mn purple; O red.} 
\label{fig:Mn84_structure}
\end{figure}

The large size of these SMMs, with numerous interacting spins, poses interesting challenges to our theoretical understanding. For example, the ground-state spin of \ch{\{Mn84\}} is quite small, $S\sim 6$, yet there is not an obvious mechanism for how the small non-zero spin arises from the interaction of 84 spins. Further, the 
spin configuration space in the \ch{\{Mn84\}} torus is $5^{84}\approx 5\times 10^{58}$, exceeding the memory of the largest supercomputers by many orders of magnitude.
In previous work, we showed that using first-principles density functional theory (DFT) derived exchange couplings, together with a coarse-grained theoretical treatment of the spins, it was possible to uncover the origin of the small but non-zero ground-state spin. In particular, the small long-range exchange couplings renormalize into coarse-grained quadratic spin-spin effective couplings between clusters of spins, generating the small non-zero ground-state spin~\cite{schurkus2020exploring}.

In the current work, we extend our previous investigation to report on
the theoretical modeling of two directly measurable experimental quantities, namely, the (temperature-dependent) field-dependent magnetization and the heat capacity of the \ch{\{Mn84\}} torus. We also experimentally measure these quantities, allowing for a direct comparison between theory and experiment in this system. 
The trajectory of the magnetization with increasing field as it transitions through successive states from $S = 6$ to the saturation magnetization 
of 344 $\mu_B$ provides information that constrains the energy scales of the interactions. Here we access part of that phase diagram in high magnetic fields up to 75 T. Similarly, features in the heat capacity vs temperature give insight into the scale of energy gaps between spin states.

Theoretically, our work also demonstrates an \emph{exact} computation of the partition function of 84 $S=2$ classical spins (in the $5^{84}$ dimensional state space). This is made possible by the adaptation of recent exact tensor network contraction methods used to simulate quantum supremacy circuits~\cite{gray2021hyper}, and demonstrates the power of such techniques beyond quantum circuit settings. We also carry out more extensive first-principles DFT calculations to estimate the long-range exchange couplings.
Our work illustrates the state-of-the-art of modeling for the physical properties of the most complicated molecular magnets.

\begin{figure}
\includegraphics[width=0.8\columnwidth]{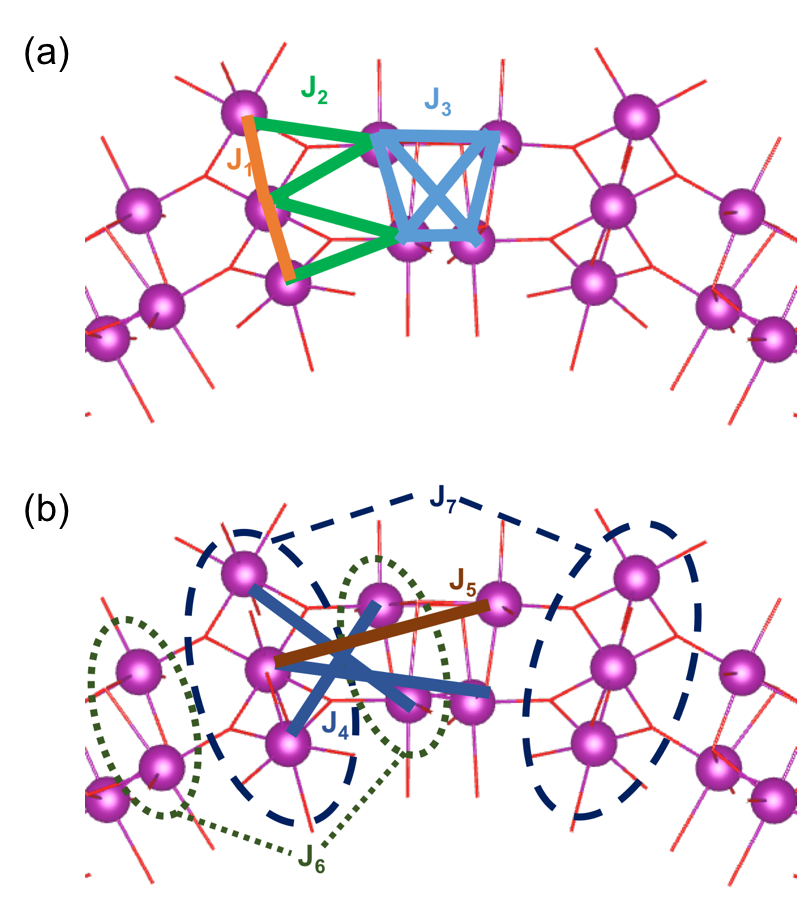}
\caption{The schematic diagram of the exchange interactions between the \ch{Mn^{3+}} in \ch{\{Mn84\}}, including (a) nearest-neighbor interactions $J_1$, $J_2$ and $J_3$ and (b) the long-range interactions $J_4$, $J_5$, $J_6$ and $J_7$.}
\label{fig:J_diagram}
\end{figure}

\begin{figure}
\includegraphics[width=0.9\columnwidth]{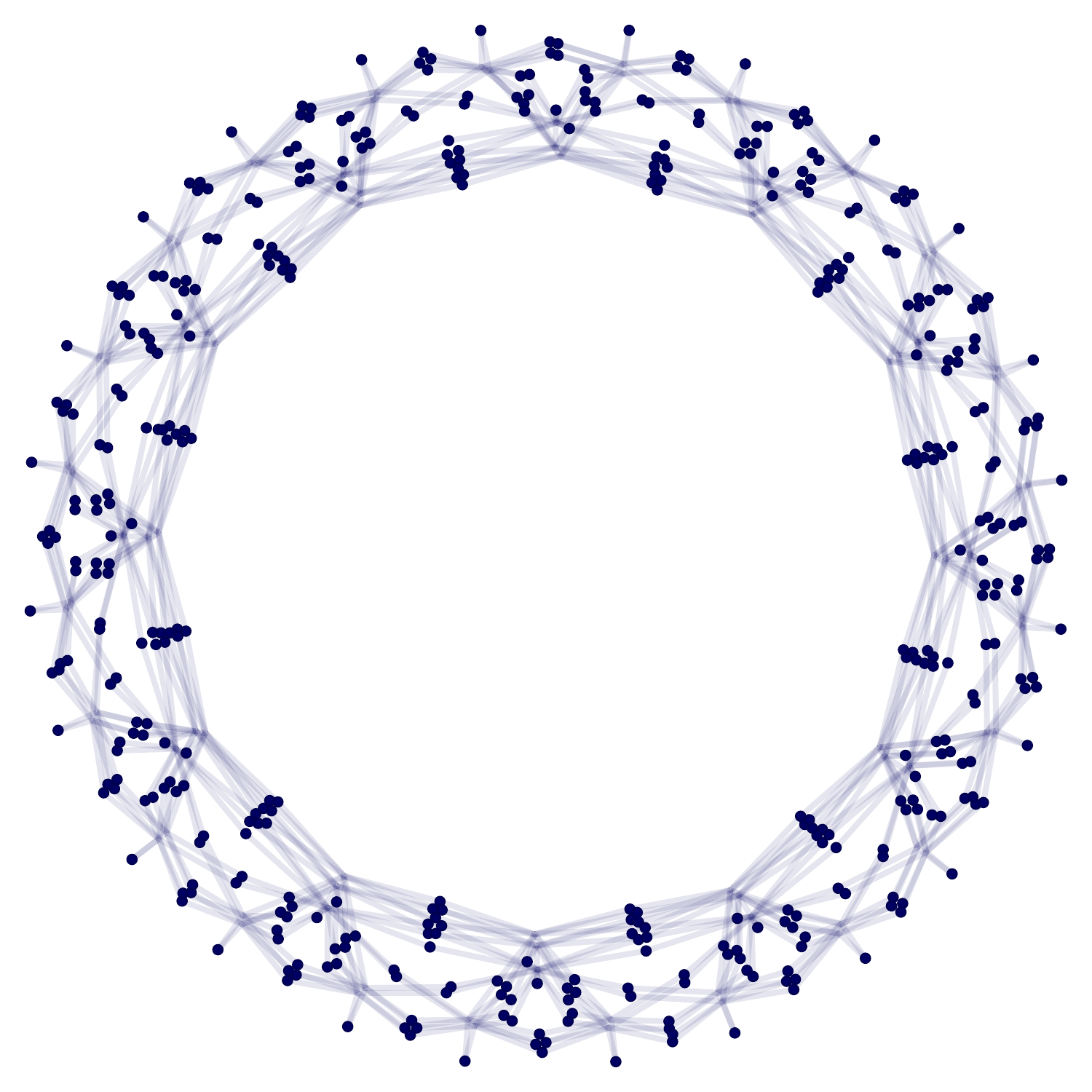}
\caption{The tensor network representation of the partition function for the classical spin model Hamiltonian.}
\label{fig:Mn84_tn}
\end{figure}

\begin{figure}[h]
\includegraphics[width=0.9\columnwidth]{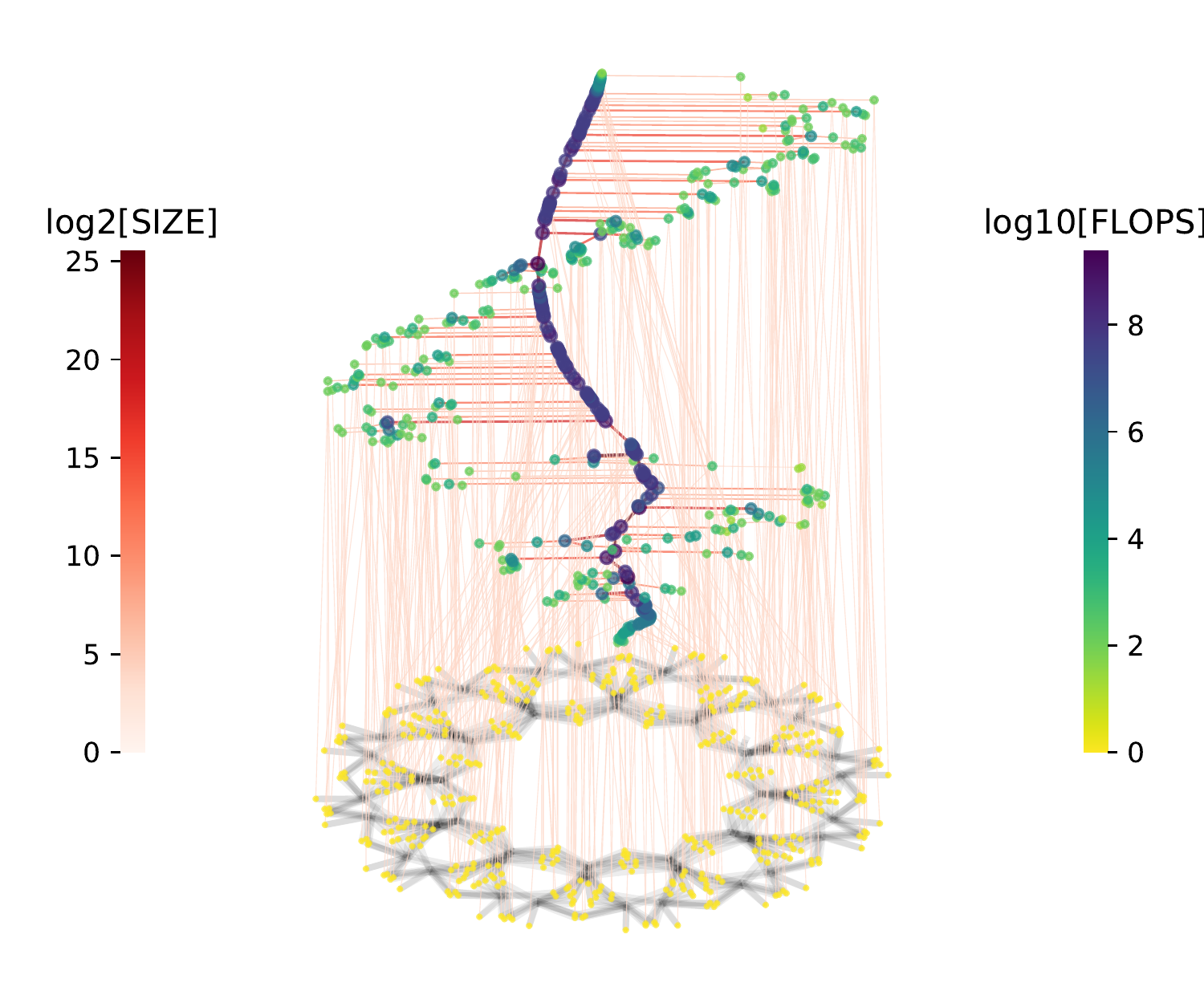}
\caption{An example contraction tree for the exact contraction of the TN representation of the canonical partition function. The partition function TN is shown on the bottom with yellow transfer matrices and gray indices. Each of the other nodes represents a contraction between tensors, with the color indicating the number of FLOPS required for the contraction. The colored lines indicate which tensors are being contracted, with the color representative of the tensor sizes.}
\label{fig:tn_contraction}
\end{figure}

\section{Theoretical and experimental methods}
\label{section:methods}

\begin{table}
\centering
\begin{tabular}{|c|c|c|c|c|c|c|} 
 \hline
  & E (DFT) & E (7J) & E (3J) &Total Spin \\
 \hline
 GS1 & 0  &0  & 0 & 24 \\
 \hline
 GS2 & 60.6 & 60.6 & 0 & 0 \\
 \hline
 GS3 & 90.5 & 91.5 & 0 &24 \\
 \hline
 GS4 & 153.7&  153.7 & 0 &0 \\
 \hline
 GS5  & 157.1&  155.1 & 0 & 12 \\
 \hline
 GS6 &  217.6&  218.6 & 0 & 0 \\
 \hline
 RS1  & 608.4  &  573.5 & 496.9 & 4\\
 \hline
 RS2  & 768.2 & 560.0 & 519.9 & 8\\
 \hline
  RS3 & 784.8 &  768.0 & 654.0  &12 \\
 \hline
  RS4 & 800.3 & 712.3 & 601.8 &4\\
 \hline
  RS5 & 919.9 & 777.2 & 692.9 &8\\
 \hline
   RS6 & 964.1 &  787.8 & 800.8 &20 \\
 \hline
 FM & 4343.7 & 4610.0 & 4230.0 & 168 \\
   \hline
\end{tabular}
\caption{Spin configurations of \ch{\{Mn84\}}. RS1 to RS6 are random spin configurations that are used to extract the nearest-neighbor interactions $J_1$, $J_2$ and $J_3$. GS1 to GS6 are selected degenerate ground state configurations of the 3J Heisenberg model. The FM configuration has all  spins aligned in the same direction (all up or all down). Energies in \meV.}
\label{table:Mn84_configs}
\end{table}

\begin{table}
\centering
\begin{tabular}{|c|c|c|c|c|c|c|} 
 \hline
$J_{1}$ & $J_{2}$ & $J_{3}$ & $J_{4}$ & $J_{5}$ & $J_{6}$ & $J_{7}$  \\
 \hline
 -13.1 & -3.3 & -1.2 &      &     &      &     \\
 -13.0 & -3.2 & -1.1 & -0.4 & 1.7 & -0.5 & -0.8\\ 
\hline
\end{tabular}
\caption{Exchange coupling constants of \ch{\{Mn84\}} obtained by fitting to DFT energies, for a 3J model (top line) and seven-J model (bottom line). Positive values are ferromagnetic couplings and negative values are antiferromagnetic couplings. Units of \meV.} \label{table:Mn84_js}
\end{table}

\begin{figure}
\includegraphics[width=1\columnwidth]{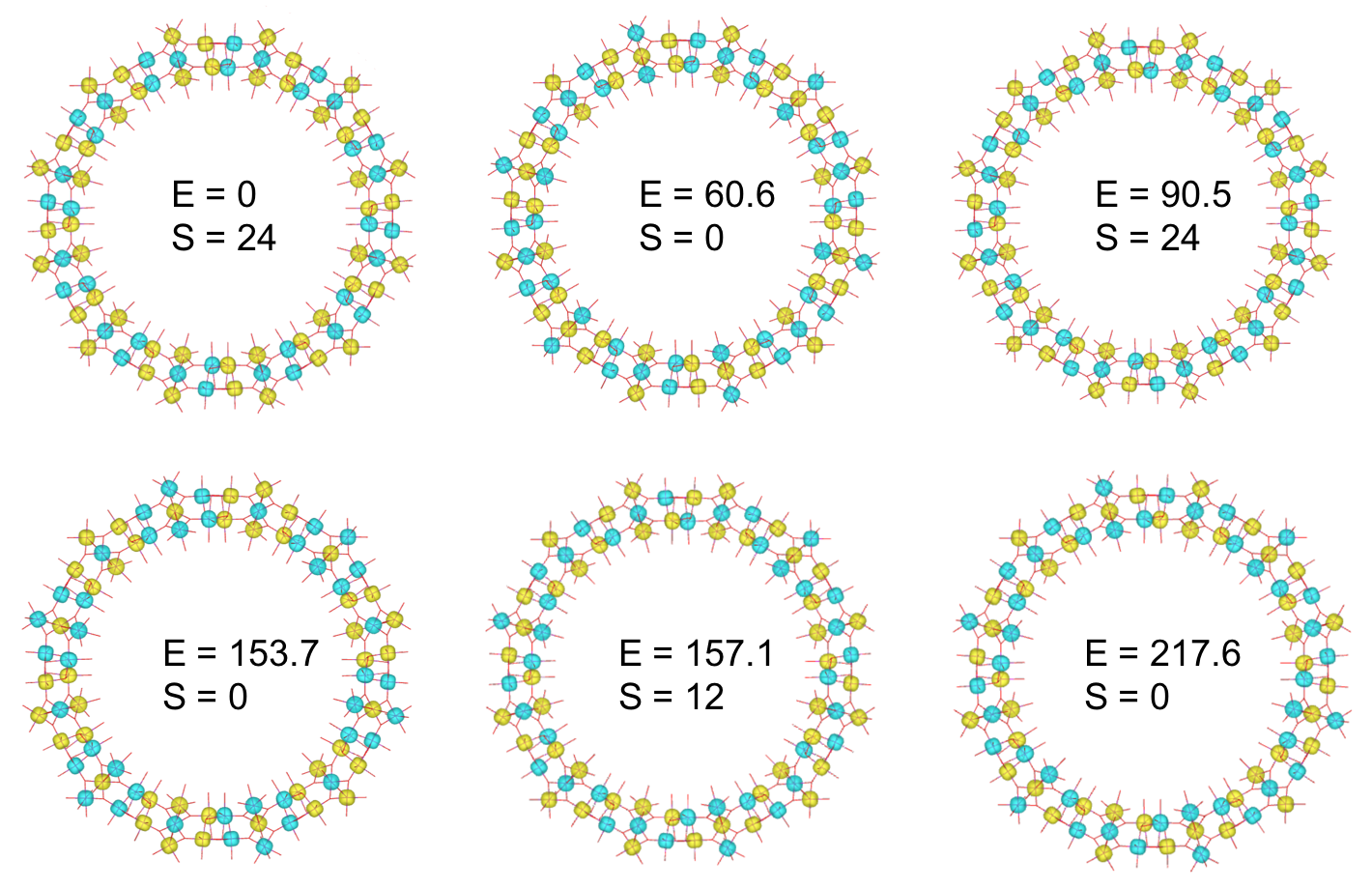}
\caption{Spin configurations of GS1 to GS6 in \ref{table:Mn84_configs}). Yellow for positive (spin up) and
  cyan for negative (spin down). Each \ch{Mn^{3+}} has a spin of $S=2$ or $S=-2$. The
  numbers in the center of each spin configuration are the total energy E in meV
  (the lowest one among the six is set to zero) and the total $S_z$.}
\label{fig:Spin_dis}
\end{figure}

\subsection{First-principles determination of exchange interactions}
\label{subsection:DFT}
We calculated electronic energies of the \ch{\{Mn84\}} torus using Kohn-Sham density functional theory (DFT) \cite{kohn1965self} with the spin-polarized Perdew-Burke-Ernzerhof (PBE) exchange correlation functional \cite{perdew1996generalized} and projector-augmented-wave (PAW) pseudopotentials \cite{blochl1994projector,kresse1999ultrasoft} in conjunction with a plane-wave basis (500~eV cutoff energy, energy convergence threshold of $10^{-6}$~eV) as implemented in the Vienna Ab-initio Simulation Package (VASP) \cite{kresse1996efficiency,kresse1996efficient}. 
We first performed ionic relaxation to obtain the optimized structure of \ch{\{Mn84\}} until the atomic forces were less than 0.05 eV/\angstrom.
Then we used a fixed optimized structure to calculate the total energies of different collinear Mn spin configurations. 

The DFT energies were fit to a Heisenberg Hamiltonian of the form
\begin{equation}
    H_{\text{Heis}}=-\sum_{i<j}J_{ij}\Vec{S_i}\cdot\Vec{S_j}
\label{eq:H}
\end{equation}
where $\Vec{S_i}$ and $\Vec{S_j}$ are $S=2$ spins (\ch{Mn^{3+}}) and $J_{ij}$ is their exchange coupling constant. We considered two different Heisenberg exchange models; a short-range 3J model with 3 nearest-neighbour interactions (Fig.~\ref{fig:J_diagram}a) and a long-range 7J model with 7 
exchange coupling interactions (Fig.~\ref{fig:J_diagram}b).

\subsection{Tensor network partition function calculations}
\label{subsection:tn_methods}

To obtain the finite-temperature field dependent magnetization and heat capacity, we computed the canonical partition function using tensor network methods. We first made the simplification that the 
spins are classical spins (which is expected to be qualitatively reasonable from the $1/S$ expansion given the relatively large Mn $S=2$ spin). We next represented the thermal properties using the 5-state Potts model,
where each spin is one of 5 integer values with $S_z\in[-2, -1, 0, 1, 2]$.
The energy of the spin configuration in the Potts model is obtained by replacing  the vector spins of the Heisenberg model with the integer spins of the Potts model, i.e.
\begin{align}
E_\text{Potts} = -\sum_{i<j} J_{ij} S_i S_j.
\end{align}
Additionally, to understand the effects of the multiple $m_s$ levels, we performed analogous calculations using an Ising model with $S_z\in[-2,2]$ and a 3-state Potts model with $S_z\in[-2,0,2]$.

Even with the above simplifications, the naive computation of the partition function requires a sum over $5^{84}$ spin configurations
\begin{align}
Z = \sum_{\{ S_i \in (-2, \ldots, 2) \}} e^{-\beta E(S_1, S_2, \ldots, S_{84})}
\end{align}
where $\beta$ is inverse temperature and $E(S_1, S_2, \ldots S_{84})$ is the energy of the classical spin configuration. To evaluate $Z$, we first re-express the summation exactly as a nested sum of products (a tensor network). This is obtained by using the Boltzmann weights of each pair of spins
\begin{align}\label{eq:tm}
Z = \sum_{\{ S_i \in (-2, \ldots, 2) \}} \prod_{i<j} e^{-\beta J_{ij} S_i S_j}
\end{align}
We can visualize the above summation structure as a graph (Fig.~\ref{fig:Mn84_tn}) where the nodes represent 
the Boltzmann weights for a pair of spins from Eq.~\ref{eq:tm};
each node has two edges, representing the spin indices $S_i$.


The cost of performing the exact summation (contraction) of such a tensor network is highly dependent on the exact order (contraction path) in which the summation is performed, with most contraction paths being prohibitive in terms of memory or computation time. A similar problem arises in the classical simulation of quantum circuits, which corresponds to a type of large tensor network contraction. Many strategies have been proposed to minimize the cost of such a tensor network contraction by finding the optimal contraction path.
To do the contraction efficiently, we use the software package \texttt{quimb}~\cite{gray2018quimb} to construct the tensor network and \texttt{cotengra}~\cite{gray2021hyper} to optimize over contraction paths while limiting memory usage and computation cost.
The resulting ``hyper-optimized'' tensor network contraction path is visualized in  Fig.~\ref{fig:tn_contraction}.
To avoid overflow errors encountered during the contraction, the logarithm of the partition function is calculated, with intermediates being stripped of exponentially large or small constants and accounted for after the full contraction. Using this strategy, the exact partition function can be computed in a few minutes on a single CPU cluster node.

Given the partition function $Z$ (and thus free energy $F = -\frac{1}{\beta} \log Z$) we evaluate the magnetization $M$ and heat capacity $C_v$ as the  derivatives
\begin{align}
M &= -\frac{\partial F}{\partial B} \\
C_v &= -\frac{1}{kT^2}\frac{\partial^2 \beta F}{\partial\beta^2},
\end{align}
computed numerically with finite difference approximations, where $k$ is Boltzmann's constant, $T$ is temperature, and the magnetic field dependence enters the energy as
\begin{equation}
    E(B, \{ S_i \} )=E_\text{Potts}-\sum_ig\mu_B{B}{S_i}
\label{eq:potts}
\end{equation}
where $g=2.0023$ (atomic units) and  $\mu_B$ is the Bohr magneton, and $B$ is the magnetic field (along the $z$ direction). To isolate the spin contribution to the heat capacity, we also compute the harmonic vibrational contribution to the heat capacity using the GFN-FF force field~\cite{spicher2020robust} to carry out a phonon calculation starting from the optimized molecular structures previously obtained by DFT.


\subsection{Experimental preparation and measurements}

Crystalline samples of \ch{\{Mn84Pr\}_{MeOH}} were prepared following the published procedure \cite{tasiopoulos2004giant}, with the modification that \ch{[Mn12O12(O2CEt)12(H2O)4]} and \ch{EtCO2H} were used as the starting materials,  and crystals were isolated after a couple weeks from layering with nitromethane.
%

Magnetization in millisecond-scale pulsed magnetic fields up to 60 T was measured on powder samples using the National High Magnetic Field Lab's standard approach to pulsed-field magnetization measurements. The measurement coil is a radially-compensated coil wound from 50 gauge copper wire. The powder samples were inserted into a non-magnetic ampule and secured with grease. The coil signal in pulsed fields is proportional to the change of magnetization in time and the signal is numerically integrated to obtain the magnetization. For each magnetization vs field curve, the ampule in and out of the coil signals under identical conditions were collected and the ampule out signal was subtracted from the ampule in signal to remove background signals. Pulsed-Field magnetization data were calibrated with the magnetization data obtained in a Vibrating Sample Magnetometer in a 14 T Physical Properties Measurement Systems (Quantum Design). The pulsed field were provided by a capacitor-driven 75~T duplex magnet at the National High Magnetic Field Laboratory in Los Alamos.

\section{Results and Discussions}

\begin{figure*}
\includegraphics[width=1\textwidth]{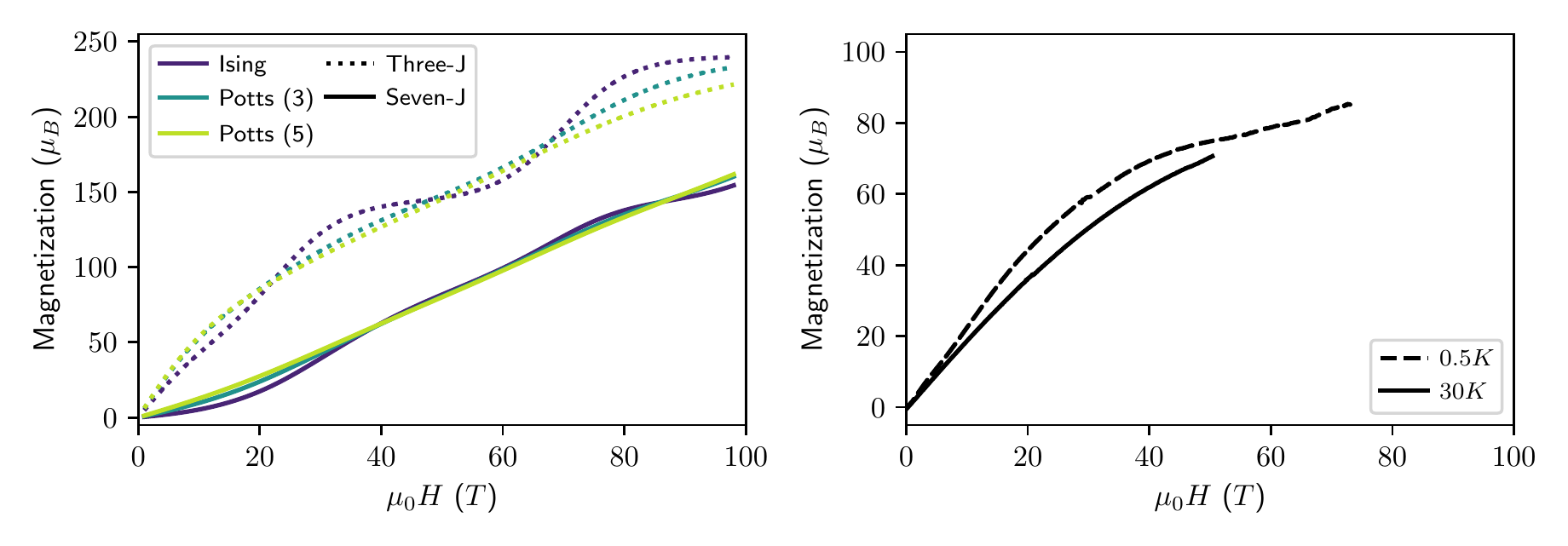}
\caption{The magnetization of \ch{\{Mn84\}} as a function of the applied field. The left panel shows the theoretical prediction using the Ising and 3- and 5-state Potts models and the 7J and 3J interactions at 30K, computed using hyper-optimized tensor network contractions. The right panel shows the experimental magnetization curves at 0.5K and 30K. $H$ is the free applied field and $\mu_0$ is magnetic permittivity of free space with $B=\mu_0H$.}
\label{fig:mag_30K}
\end{figure*}

\begin{figure}
\includegraphics[width=1\columnwidth]{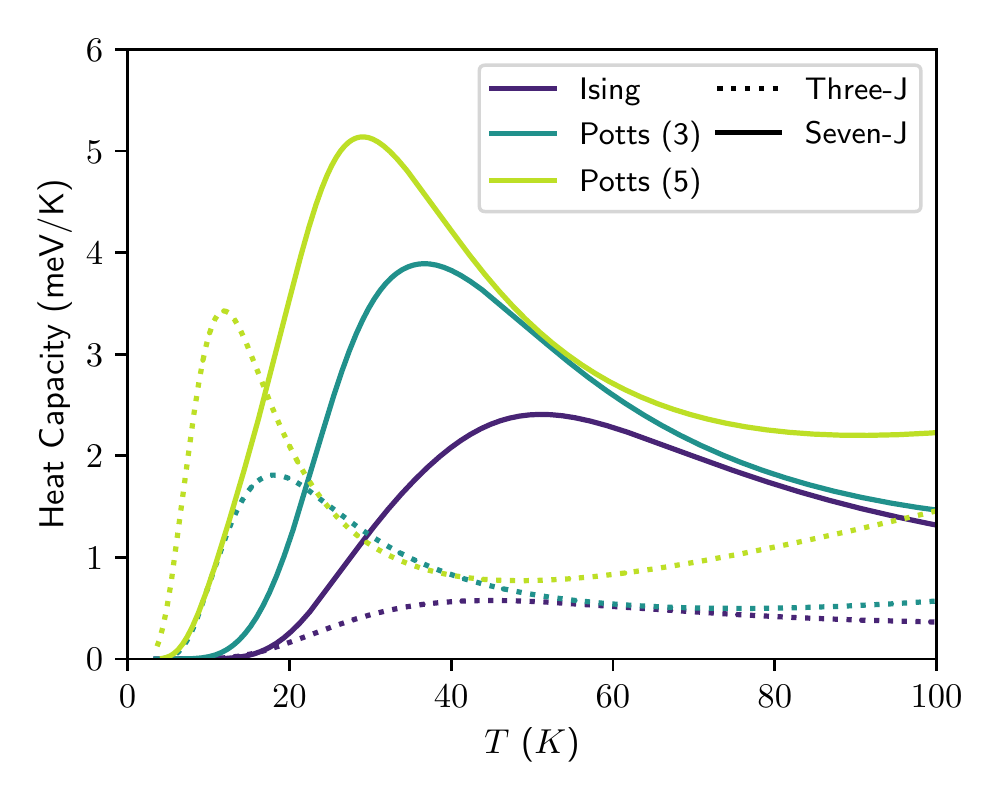}
\caption{Calculated spin contribution to the heat capacity of \ch{\{Mn84\}} with the Ising and 3- and 5-state Potts models, using the 7J and 3J interactions.  Results are calculated with no applied magnetic field using hyper-optimized tensor network contractions.}
\label{fig:C_v}
\end{figure}

\begin{figure*}[h]
\includegraphics[width=1\textwidth]{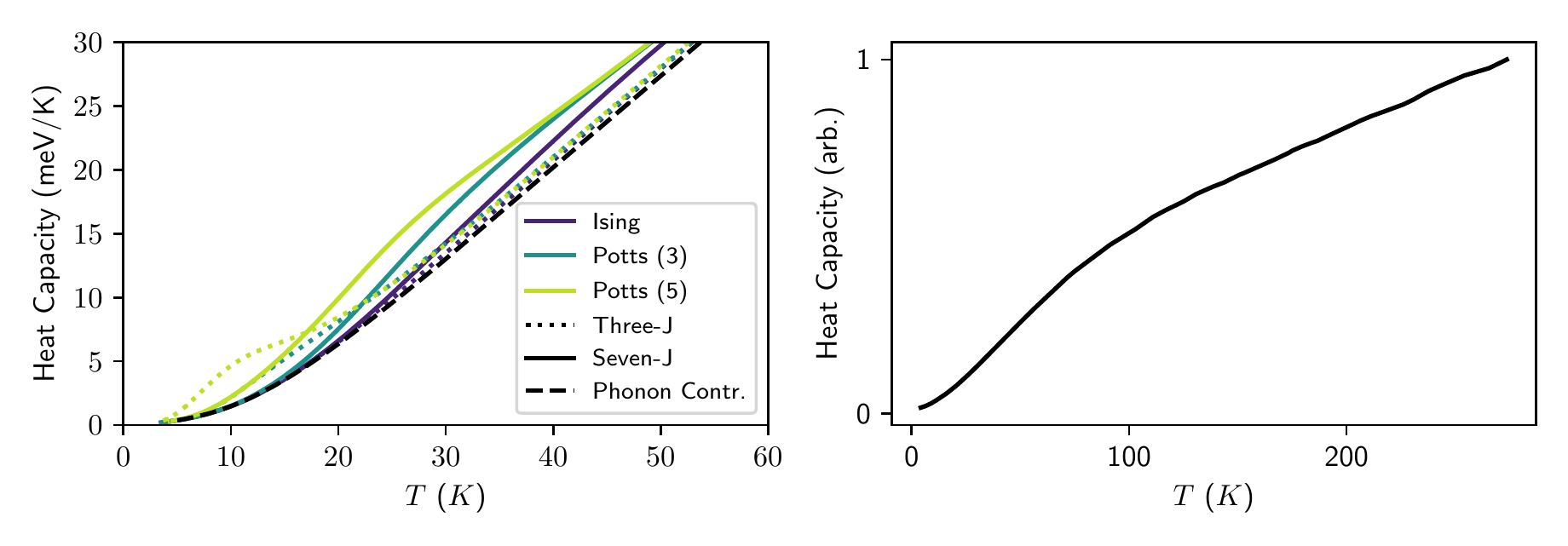}
\caption{The calculated heat capacity as a function of temperature with no magnetic field. In the left panel, the theoretical prediction is sown, with the phonon contribution to the heat capacity shown as the black dashed line, while the solid and dotted lines include the magnetic contribution, computed via the tensor network calculations for the discussed models. The right panel shows the corresponding experimental curve in arbitrary units.} 
\label{fig:heat_capacity_experiment}
\end{figure*}

\subsection{Exchange Interaction Constants}
\label{subsection:coupling}

Using the first principles procedure described above, we compute the energies of 13 different Ising-like (i.e. each Mn spin is maximally aligned along the $z$ axis, pointing out of the plane of the wheel) spin configurations of \ch{\{Mn84\}}   (Table \ref{table:Mn84_configs} and Fig.~\ref{fig:Spin_dis}). 


As discussed in~\cite{schurkus2020exploring}, using only the 3J model ($J_1$, $J_2$, $J_3$ in Fig.~\ref{fig:J_diagram}) results in a large ground-state degeneracy in a  classical Ising or Potts model, with ground-state spins ranging from $S=0$ to $S=24$.
 GS1 to GS6 in Table~\ref{table:Mn84_configs} are six of the ground state spin configurations that would be degenerate within a 3J model. 
 However, as the DFT results show, there is at least a 217.6 \meV{} energy spread between the lowest and highest energies of these "degenerate" states. Thus, the energies are not actually degenerate, presumably due to the longer range interactions, which are also the interactions that select a specific ground-state spin. 

 To determine the 3J parameters, as well as to estimate the possible
 long-range exchange couplings, we consider the additional configurations RS1-RS6 (randomly chosen) as well as the ferromagnetic configuration FM.
 %
 We construct a best fit of the 3J model to all these configurational energies.
In addition, we introduce long-range interactions, in particular, the $J_4-J_7$ couplings shown in Fig.~\ref{fig:J_diagram}, to model
the splitting of the ground-state degeneracy. 
The results of the fit to the 3J, 7J models are shown in Table~\ref{table:Mn84_configs}; the values of the 3J and 7J parameters are shown in Table~\ref{table:Mn84_js}. From Table~\ref{table:Mn84_js} we see that the long-range couplings are all small and alternate in sign, while 
Table~\ref{table:Mn84_configs} shows that there are still substantial errors in this fit (see e.g. the reversal of the order of RS3 and RS4) reflecting the difficulty in both fitting and perhaps the limitations of the spin Hamiltonian itself. Nonetheless, the main trends are reproduced. We assess the quality of the exchange couplings against experiment in the next section.

\subsection{Magnetic susceptibility and heat capacity}

As described in Sec.~\ref{subsection:tn_methods}, 
starting from an Ising or Potts Hamiltonian using the 3J and 7J models obtained above, we used hyper-optimized tensor network contraction to calculate the magnetic susceptibility and heat capacity. We now compare the results of this parameter-free theoretical treatment to the experimental measurements.


The theoretical magnetization at 30K is shown in the left panel of Fig.~\ref{fig:mag_30K}. 
The Ising model curves show a clear inflection point occurring respectively for the 3J and 7J models around 150$\mu_B$ and 75$\mu_B$ at magnetization strengths between 30-70T. The large difference in the vertical magnitude of these curves, however, illustrates the sensitivity to the accuracy of weak long-range exchange couplings. The inflection is reduced as the number of local states increases from the two states of the Ising model to the 5-state Potts model.

Experimental magnetization measurements are shown in the right panel of Fig.~\ref{fig:mag_30K} at 0.5K and 30K. 
While the lower temperature curve shows an inflection similar to the Ising model result, the higher temperature experimental data mirrors the behavior of the Potts models. The experimental magnetization is somewhat smaller in magnitude than results calculated with the 7J models and significantly smaller than the 3J predictions. Beyond inaccuracies in the long-range exchange couplings highlighted above, another source of discrepancies is the form of the magnetic Hamiltonian: the reduction from the Heisenberg to Potts models, neglect of magnetic anisotropy, as well as the alignment of molecules relative to the field in the experimental sample.

The theoretical heat capacity from the spin-degrees of freedom is shown in Fig.~\ref{fig:C_v}, while the total heat capacity (i.e. including the contribution of the phonons) is shown together with the experimental heat capacity in Fig.~\ref{fig:heat_capacity_experiment}, both in the absence of a magnetic field.
For all models, there are two peaks: a sharp one at low temperatures ($\sim 15-60$K) and a broad one at higher temperatures ($\sim 200$K and above, not shown in figure). 
The precise temperature at which the peak occurs is strongly model-dependent.
The large spin contribution to the heat capacity manifests as a bump in the total heat capacity when superposed on top of the background phonon contribution. As seen in the first inset, the experimental measured heat capacity shows a bump at around a temperature of nearly 100K. The bump in the theoretical total heat capacity comes from the first peak in the spin heat capacity, anywhere between 10-50K depending on the model.

\section{Summary}

\label{section:conclusion}

In the present work we carried out theoretical simulations of the giant \ch{\{Mn84\}} wheel directly targeting two of the common experimental observables used to characterize single molecule magnets,  the heat capacity and field-dependent magnetic susceptibility. Our theoretical simulations contained no adjustable parameters and featured both large-scale first principles calculations of the exchange interactions as well as new hyper-optimized tensor contraction methods for the partition function.

The most interesting features in the observables for \ch{\{Mn84\}} were the  peaks in the heat capacity and the inflection in the field-dependent magnetization, seen in both theory and experiment. Both must arise from the degeneracy structures we identified previously in the \ch{\{Mn84\}} energy spectrum. 
The computed and experimental curves for the heat capacity and magnetization resemble each other, although the energy scales in the computation are shifted from experiment, and appear very sensitive to fine details in the Hamiltonian. Nonetheless, the qualitatively good agreement illustrates the increasing theoretical capabilities to model low-energy physics even in the most complicated single molecule magnets.

\bibliography{References}

\end{document}